\documentclass[aps,prb,reprint, superscriptaddress, showpacs]{revtex4-1}

\usepackage{amsmath,amsfonts,
}
\usepackage{color}
\usepackage{color, graphicx}
\usepackage{dcolumn}
\usepackage{bm}
\usepackage{amssymb}
\usepackage{amsmath}
\usepackage{latexsym}
\usepackage{amsfonts}
\usepackage{amsmath}
\usepackage{multirow}
\usepackage{ifthen}

\begin{document}

\title{Acoustic Bessel-like beam formation by an axisymmetric grating}

\author{N. Jim\'enez}
\affiliation{Instituto de Investigaci\'on para la Gesti\'on Integrada de Zonas Costeras, Universitat Polit\`ecnica de Val\`encia, Paranimf 1, 46730 Grao de Gandia, Spain}

\author{V. Romero-Garc\'ia}
\affiliation{LUNAM Universit\'e, Universit\'e du Maine, CNRS, LAUM UMR 6613, Av. O. Messiaen, 72085 Le Mans, France}

\author{R. Pic\'o}
\author{A. Cebrecos}
\author{V.J. S\'anchez-Morcillo}
\affiliation{Instituto de Investigaci\'on para la Gesti\'on Integrada de Zonas Costeras, Universitat Polit\`ecnica de Val\`encia, Paranimf 1, 46730 Grao de Gandia, Spain}

\author{L.M. Garcia-Raffi}
\affiliation{IUMPA, Universitat Polit\`ecnica de Val\`encia, Cam\'i de Vera s/n, 46022 Valencia, Spain}

\author{J.V. S\'anchez-P\'erez}
\affiliation{Centro de Tecnolog\'ias F\'isicas: Ac\'ustica, Materiales y Astrof\'isica, Universitat Polit\`ecnica de Val\`encia, Cam\'i de Vera s/n, 46022 Valencia, Spain}

\author{K. Staliunas}
\affiliation{ICREA, Departament de F\'isica i Enginyeria Nuclear, Universitat Polit\`ecnica de Catalunya, Colom, 11, E-08222 Terrassa, Barcelona, Spain}

\begin{abstract} We report Bessel-like beam formation of acoustic waves by means of an axisymmetric grating of rigid tori. The results show that the generated beam pattern is similar to that of Bessel beams,
characterized by elongated non-diffracting focal spots. A multiple foci structure is observed, due to the finite size of the lens. The dependence of the focal distance on the frequency is also discussed, on the basis of an extended grating theory. Experimental validation of acoustic Bessel-like beam formation is also reported for sound waves. The results can be generalized to wave beams of different nature, as optical or matter waves.
\end{abstract}


\maketitle

Bessel beams, originally proposed in optics \cite{Durnin87, Maddox87}, are now at the basis of many applications due to their unusual propagation properties\cite{Duocastella12, Fahrbach11, Chu12, Matsupka06, Hsu89, Katchadjian10}. The most celebrated property of a Bessel beam is that, in the ideal case, the field propagates invariantly, i.e. without any diffracting broadening, in contrast to the other canonical case, the Gaussian beam, where the beam experiences diffractive broadening in free space propagation. As a consequence, the field pattern in a Bessel beam possesses an infinitely extended focal line. 

\begin{figure}
\includegraphics[width=70mm]{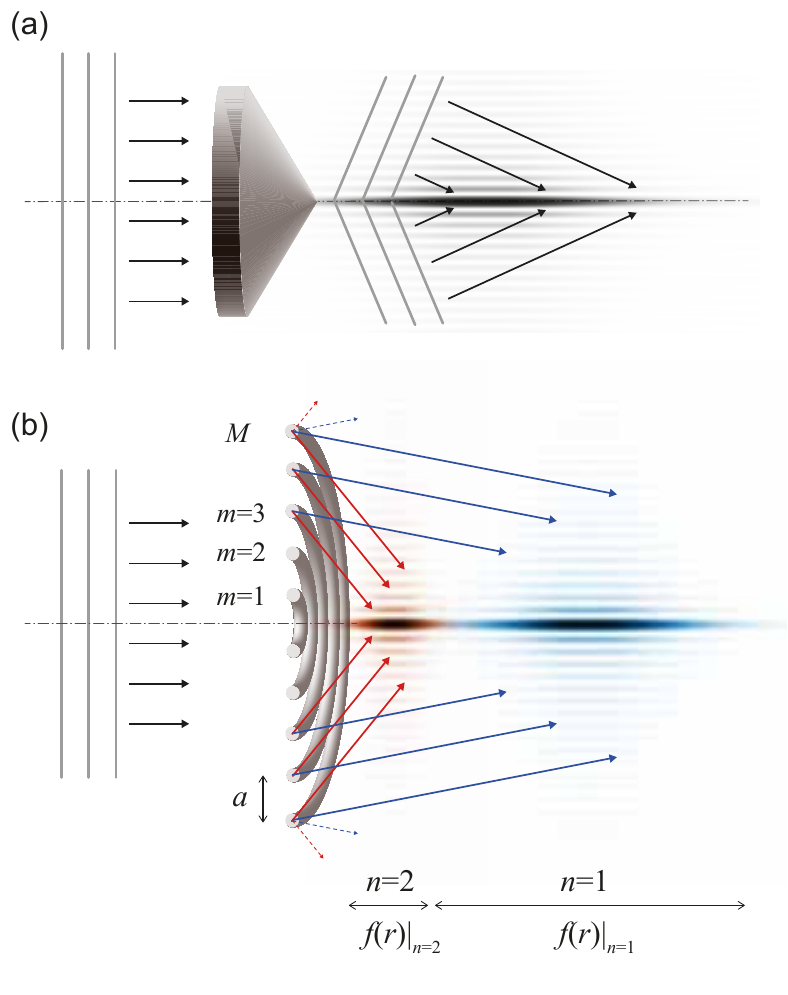}
\caption{(Color online) (a) Illustration of the formation of Bessel beam by an axicon resulting in imperfect Bessel beam showing a focus-line of finite extent; (b) Illustration of the formation of Bessel-like beams by a plane of concentric rings, where converging diffracted waves result in two elongated foci.}
\label{fig:fig1}
\end{figure}

Strictly speaking, Bessel beam is a solution of the wave equation in the form of a monochromatic wave with a transverse profile given by a Bessel function of the first kind, which by definition presents an infinite spatial extension. This ideal case cannot be realized in practice (in the same way as ideal, infinitely extended plane waves cannot exist). However, approximate or imperfect Bessel beams of finite transverse extent can be excited by different means, displaying not an infinite but an extremely elongated focal line.

In optics, Bessel-like beams are usually formed by focusing a Gaussian beam by an axicon\cite{McLeod54}, a transparent refractive element of conical shape, as shown in Fig. \ref{fig:fig1}(a). The beam in propagation through the axicon acquires linearly tilted (conical) wave-fronts, which results in an elongated focus behind the axicon. As the axicon is not infinitely extended in transverse space, the resulting Bessel beam is not perfect, and displays a focal line of finite extent. Optical Bessel beams have been also obtained by acoustic gradient index lenses \cite{McLeod06}. In electromagnetism, Bessel-like beams have been generated from a subwavelength aperture by adding a metallic circular grating structure in front of the aperture \cite{Li09}. Such imperfect Bessel beams find multiple applications, e.g. in optics for laser inscription of patterns deep into transparent materials, or for etching of deep narrow holes in laser manufacturing of opaque materials, among others \cite{Duocastella12, Matsupka06, Fahrbach11}. 

\begin{figure*}
\includegraphics[width=150mm]{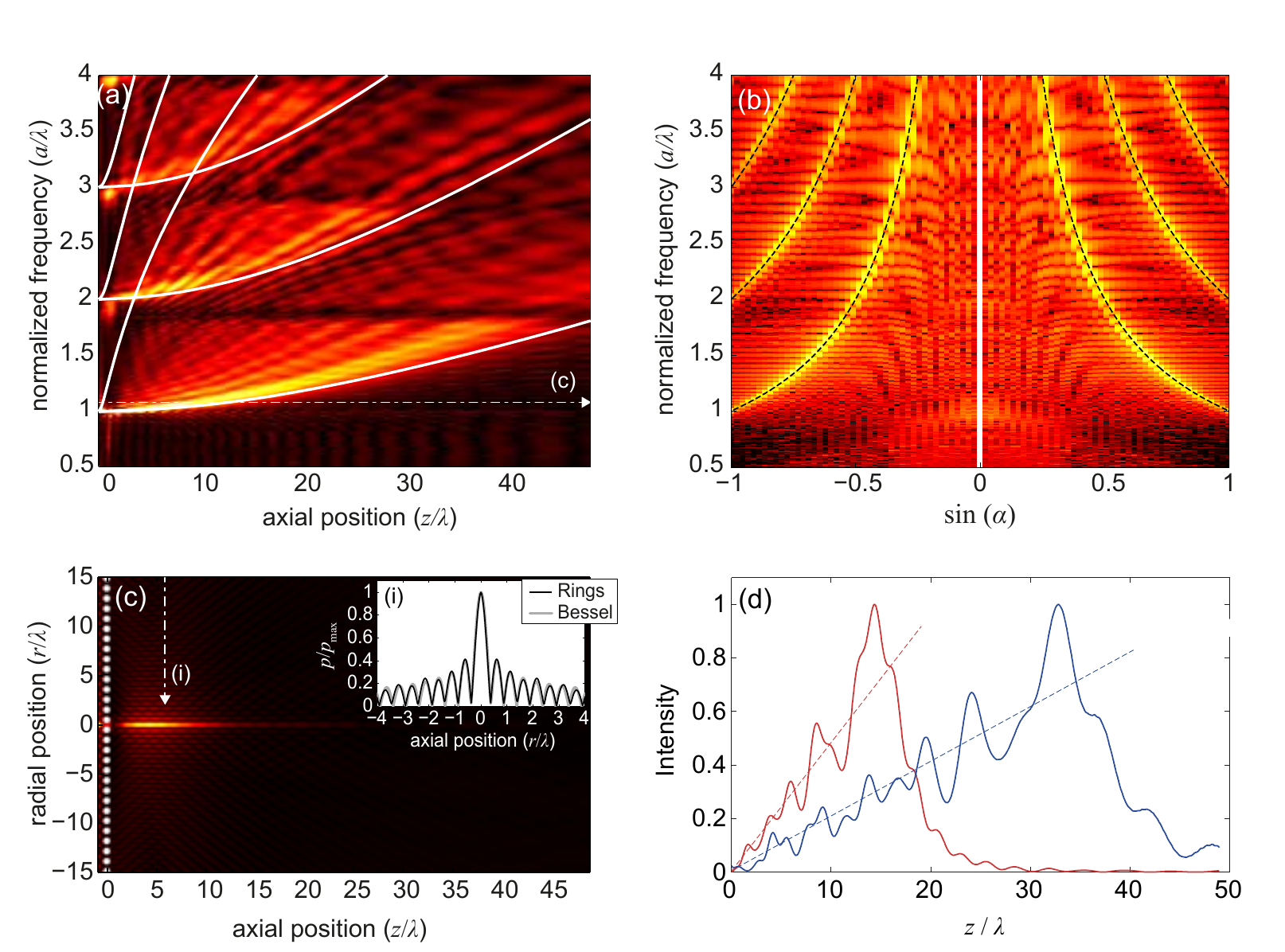}
\caption{(Color online) (a) Map of acoustic pressure along the symmetry axis (horizontal axis) for varying frequency (vertical axis). Analytical estimations from Eq. (\ref{eq:eq1}) of focal line extent from $f_{n1}$ to $f_{nM}$ are shown by solid white lines. (b) Map of far field  in angular direction (horizontal axis) for varying frequency (vertical axis), where analytical estimations are shown in dashed black lines. (c) 2D pressure distribution for case of single focus ($a = 1.03\lambda$) on a cross section along the symmetry axis. (i) Inset  shows axial cross-section of the pressure field at a distance $z = 5\lambda$ (solid black) compared with Bessel beam (solid gray). (d) On-axis intensity for the cases $a/\lambda=1.1717$ (red line) and $a/\lambda=1.6313$ (blue line). Dashed line is an eye-guide to show the linear dependence.}
\label{fig:fig2}
\end{figure*}

In acoustics, Bessel beams of sound waves were also reported \cite{Hsu89,Lu90}, however are still not so broadly applied as in optics, which is perhaps related with the lack of convenient techniques of formation of such kind of acoustic waves. Acoustic Bessel beams have been excited using acoustical axicons \cite{Katchadjian10}, in analogy to the optical case. However the most convenient way to form acoustic Bessel beams is by using annular transducer arrays \cite{Masuyama99}. Related theoretical studies include the scattering of Bessel beams by spheres \cite{Marston07}, nondiffracting bulk-acoustic $X$ waves \cite{Salo99} or non linearly generated Bessel beams of higher harmonics \cite{Ding00, Cunningham00}. 

The present work proposes and demonstrates a technique for acoustic Bessel-like beam formation using a planar structure made of of concentric tori of circular section, called here rings for simplicity. We show that, under specific conditions, part of a diffracted wave collimates, producing an elongated focus. Moreover, different diffraction orders can result in different elongated foci, as illustrated in Fig. \ref{fig:fig1}(b). In the present work we demonstrate the feasibility of this idea by analytical estimations, numerical simulations and experiments. A simple analytical model based on an approach of axisymmetric diffraction gratings is used to estimate the focal positions and the extent of the focal line. Numerical calculations (using finite difference time domain (FDTD) techniques \cite{Taflove00}) of acoustic waves propagating through such axisymmetric gratings were used to observe the complete acoustic field. Finally, the experimental verification of Bessel-like beam formation by an axisymmetric grating is reported.

Each element of the concentric ring structure is characterized by two parameters, ($i$) the toroidal radius, $r_m$ and ($ii$) the radius of the tube (circular section), $R_m$. The rings in the axisymmetric grating have increasing toroidal radii as $r_m=ma$, where $a$ is the separation between rings and $m$ is an element index, as shown in Fig. \ref{fig:fig1}(b). The continuity of the transversal component of the wave vector at the interface between the free propagation medium and a linear diffraction grating with periodicity $a$, results in diffraction of normal incident plane waves at diffraction angles given by $\sin{\alpha_n}=n\lambda/a$ where $\lambda$ is the wavelength and $n$ is the diffraction order. It is worth noting that approximately half part of the diffracted radiation converges towards the symmetry axis and the other half diverge. Resulting from the converging radiation, as it follows from simple trigonometry considerations, each ring with major radius $r_m$ is mapped to a particular distance along the symmetry axis, given by
\begin{equation}
f_n(r_m)=f_{nm}=\frac{r_ma}{n\lambda}\sqrt{1-\left(\frac{n\lambda}{a}\right)^2}.
\label{eq:eq1}
\end{equation}
If the system of concentric rings extends from $r_1$ (toroidal radius of inner ring) to $r_M$ (radius of outer ring) in the transverse plane, the focal line for the $n$-th diffraction order will extend approximately from $f_{n1}$ to $f_{nM}$. In a limiting case of infinitely extended ring structure ($r_1=0$, $r_M=\infty$) Eq. (\ref{eq:eq1}) predicts an infinitely extended focus, similarly to that of an ideal Bessel beam. 

First, we perform numerical simulations in order to explore the character of the elongated focal line (also different foci) due to the axisymmetric diffraction. For that, we use a structure composed by a set of 50 concentric rings with constant minor radius $R=a/3$, irradiated by a plane wave. We notice that this structure is much larger than that used in experiments (detailed below), therefore,  the Bessel-like features of focal line are more pronounced. For the numerical simulations we calculate the wave propagation using the FDTD technique. Fig. \ref{fig:fig2}(a) and \ref{fig:fig2}(b) represent the frequency dependence of the on-axis amplitude and radial far field amplitude respectively. Color map represents the amplitude of the acoustic field, $|p|$, and continuous lines in Fig. \ref{fig:fig2}(a) show the predictions from Eq. (\ref{eq:eq1}) for the cases $m=1$ and $m=50$ for the first  three  diffraction orders, $n=1,2,3$. The focal spots appear at normalized frequencies $a/\lambda=n$, so the diffraction angles at these frequencies corresponds to $\alpha_n=\pi/2$ ($\sin{\alpha_n}=1$). With increasing frequency, as shown in Fig. \ref{fig:fig2}(a), foci elongate corresponding to analytical estimations by Eq. (\ref{eq:eq1}) (white lines in Fig. \ref{fig:fig2}(a)). The diffraction angle decreases with the increasing frequency. Concerning the angular far field distribution shown in Fig. \ref{fig:fig2}(b), excellent agreement between theory and numerical results are observed. Different maxima appear due to different diffraction order, as a consequence of the focusing effect at near field due to the finite size of the structure, in agreement with Fig. \ref{fig:fig2}(a).

\begin{figure}
\includegraphics[width=75mm]{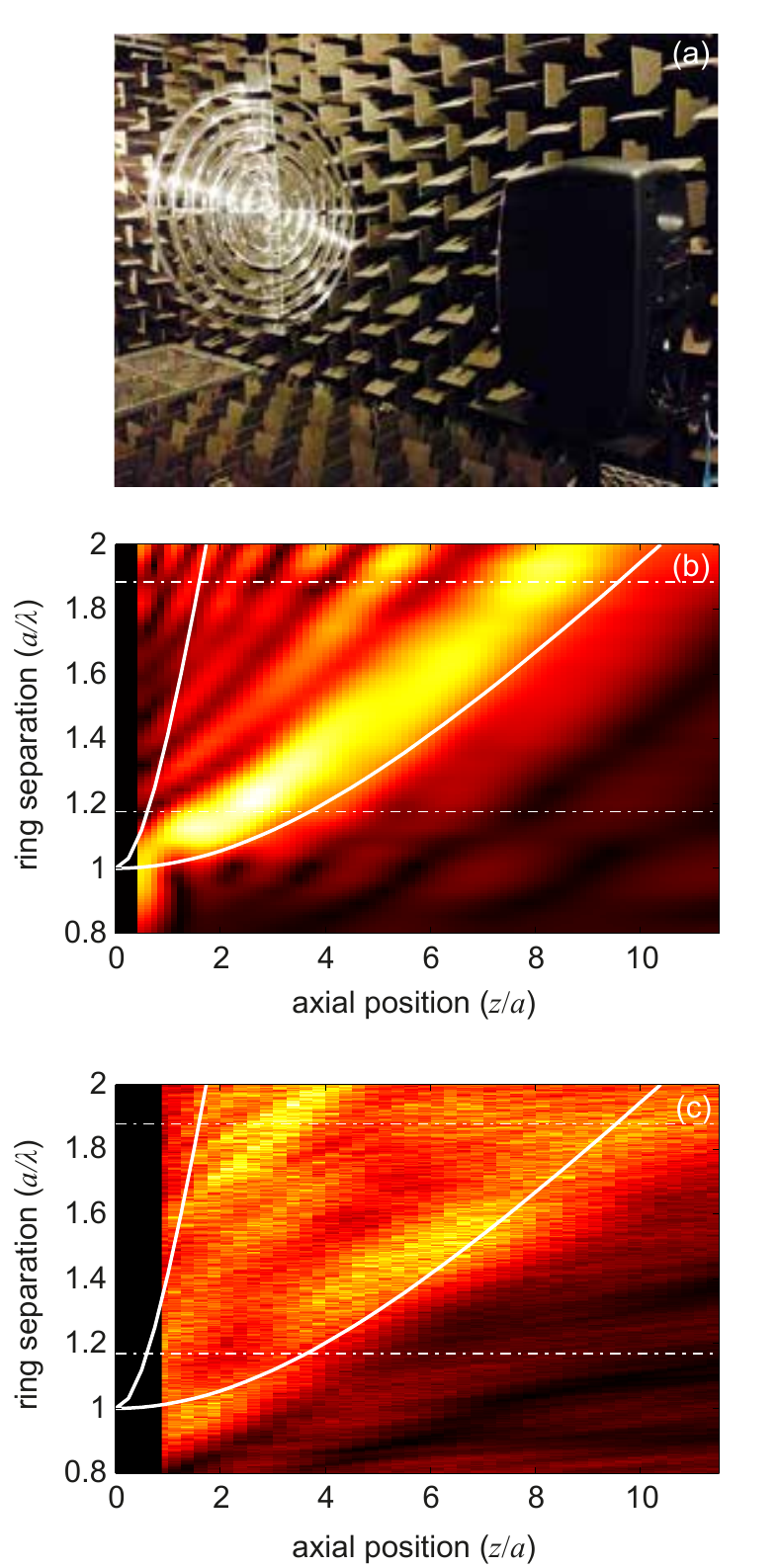}
\caption{(Color online) (a) Experimental set-up. (b) Numerically calculated and (c) experimentally measured maps of on-axis amplitude dependence for varying frequency (vertical). Solid white lines indicate the extend of the focal line from analytical estimations using Eq. (\ref{eq:eq1}) for $f_1(r_{1})$ and $f_1(r_{7})$. Dashed white lines represent the frequencies shown in Fig. \ref{fig:fig4}}
\label{fig:fig5}
\end{figure}

Next, we analyze the acoustic field behind the axisymmetric grating for a particular frequency corresponding to $a=1.033\lambda$ (see horizontal dashed line in Fig. \ref{fig:fig2}(a)). We study here the case when only one focal line appears. Fig. \ref{fig:fig2}(c) shows the field distribution on axial cross-section. The focus is substantially long, which is a signature of a Bessel-like beam. The latter is proven and illustrated in the inset of Fig. \ref{fig:fig2}(c), where the transverse field profiles at the indicated distance behind the ring structure is plotted and compared with Bessel function. 

Finally, Fig. \ref{fig:fig2}(d) shows the intensity distribution along the symmetry axis for two different frequencies: important to note that the amplitude of the field increases with distance until a maximum focusing distance, where the amplitude drops. The scattered energy is proportional to the area between neighbouring rings. Henceforth, the acoustic intensity along the focus increases with a linear trend overlaid of stray oscillations from $f_{n1}$ to $f_{nM}$, as shown in Fig. \ref{fig:fig2}(d). These oscillations are mainly due to the finite size of the structure: the edges of the diffraction grating result in fringes. We also note, that although the focal lines associated with different diffraction orders partially overlap, the radiation due to higher order diffraction clearly dominates in the interference picture. The linear dependence of acoustic intensity along the focus is however modified if the thickness of the rings depends on their radius (see the experiments below).

\begin{figure*}
\includegraphics[width=150mm]{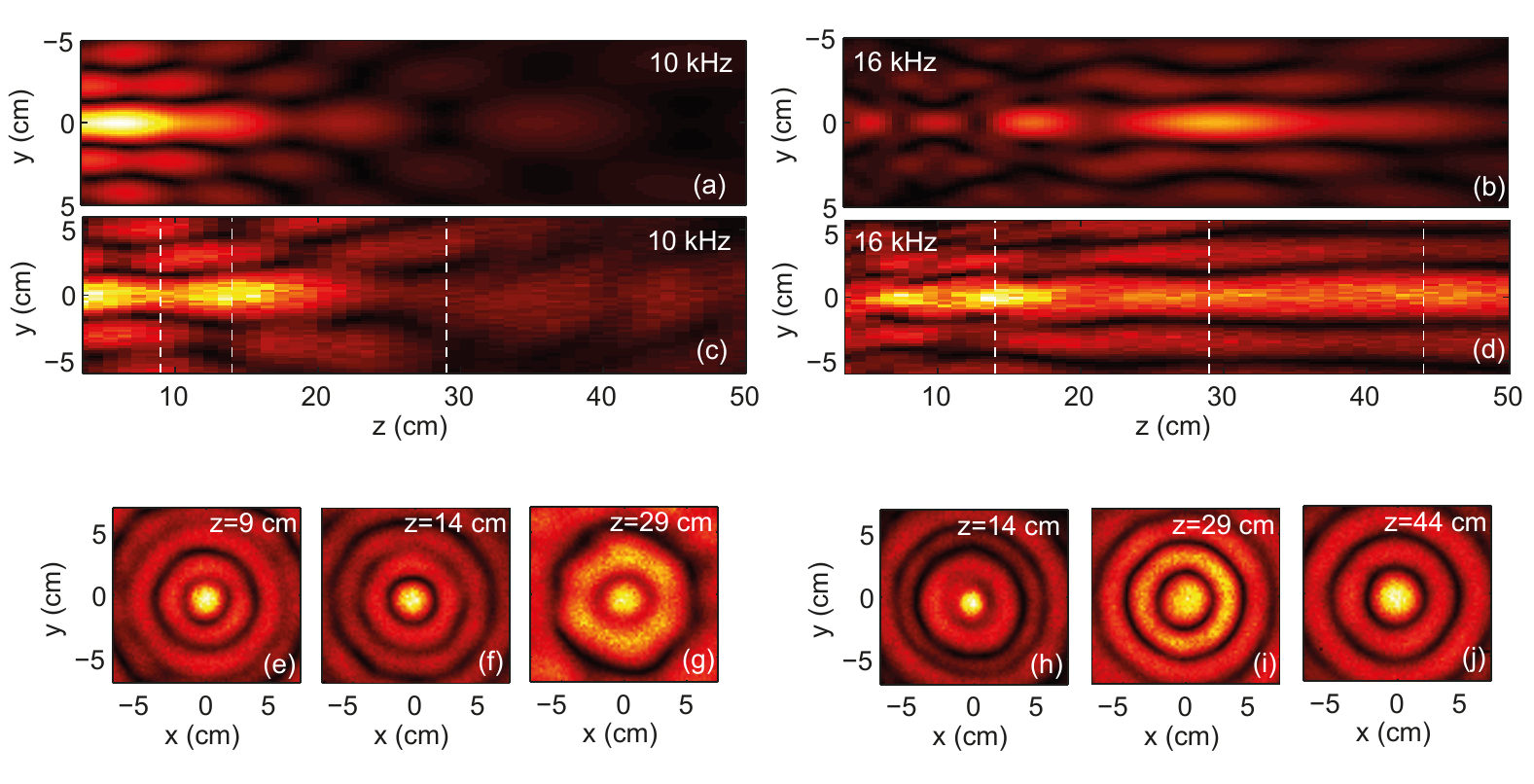}
\caption{(Color online) (a-d) Amplitude  distributions in planes along the symmetry axis $r(z, x)$ obtained by numerical simulations (a, b) and experimental (c, d) measurements for two different frequencies, showing the formation of elongated foci. Experimental transversal amplitude cross-sections (x, y) at different distances behind the ring structure, showing the formation of the Bessel-like beam. (e-g) correspond to the experimental transversal planes at 10 kHz marked with dashed lines in (c). (h-j) correspond to the experimental transversal planes at 16 kHz marked with dashed lines in (d).}
\label{fig:fig4}
\end{figure*}

For the experimental validation we designed a system composed by a set of 7 concentric rigid rings embedded in air. The rings are made of methyl methacrylate (plexiglass), which acoustic impedance is much larger than that of the air ($Z_{plexiglass}/Z_{air}\sim6000$); therefore, they can be considered acoustically rigid. The variation of the toroidal radius in the structure is $r_m=ma$, where $a=4$ cm.  The minor radius now is a function of the major radius, $R(r_m)$, following a hyperbolic secant profile like in the axisymmetric gradient index lens analyzed in Ref. [\onlinecite{Romero14}]. This profile modifies the linear dependence of the intensity along the focus, however also eliminates the sharp drop of intensity at the end of the focal line, which results in a more smooth and better reproducible diffraction pattern. All measurements were performed in anechoic chamber to avoid unwanted reflections. The source was a loudspeaker radiating a sound wave with a white noise spectrum placed at 1.5 m in front of the plane of concentric rings, a sufficient distance in order to ensure that a nearly plane wave radiates the structure at the frequencies of interest. A movable microphone located behind the structure recorded the transmission spectrum. The experimental set-up is shown in Fig. \ref{fig:fig5}(a).

The quantitative study of Bessel-like beam formation is summarized in Figs. \ref{fig:fig5}(b) and \ref{fig:fig5}(c). Pressure color maps of Figs. \ref{fig:fig5}(b) and \ref{fig:fig5}(c) show the numerical and experimental frequency dependence of the on-axis amplitude produced by the used structure. 
In the experiment, we were able to measure frequencies until 20 kHz ($\frac{a}{\lambda} <2.3$), i.e. we could achieve Bessel beam formation by the first diffraction order only. The expected dependence of the focal distance on frequency (compare with Fig. \ref{fig:fig2}) is also evident in both numerical and experimental plots.

We have also measured field cross-sections: an horizontal plane containing the symmetry axis (see Figs. \ref{fig:fig4} (a-d)) as well as several axial cross-sections (see Figs. \ref{fig:fig4} (e-j)).
We focused on two particular frequencies, 10 kHz ($a/\lambda=1.17$) and 16 kHz ($a/\lambda=1.88$) (indicated by white dashed-dotted horizontal lines in Figs. \ref{fig:fig5}(b) and \ref{fig:fig5}(c)). 
For the analyzed structure the frequency of 10 kHz is close to the condition $a/\lambda=1$, consequently, the first diffraction spot appears just behind the structure (see white dashed line in Fig. \ref{fig:fig5}(c) as a reference). However, the frequency of 16 kHz produces the elongated focus or Bessel-like beam. These two phenomena are both numerically and experimentally shown in Figs. \ref{fig:fig4}(a)-(c) and Figs. \ref{fig:fig4}(b)-(d) with good agreement.

In order to see the symmetry quality of the beams produced experimentally, we have also measured the axial cross-sections of the pressure field at different $z$-positions for above discussed cases of 10 kHz and 16 kHz (Fig. \ref{fig:fig4}(e)-(j)). 
In both cases ($i$) the diffracted pattern is highly axisymmetric and ($ii$) the diffracting broadening of the central beam, along the extended focus, is almost negligible. The transversal profiles in Figs. \ref{fig:fig4}(e)-(j) also illustrate the typical shape of the truncated Bessel function. It is simple to predict the tendencies of the amplitude distribution along the focus for such a small number of rings. The longitudinal shape along the elongated focus seems to be not linearly increasing/sharply dropping, but correspondingly smoother. This is due to the small number of rings in the structure; also due to the difference in thickness of the rings.

Concluding, we have demonstrated the principle of Bessel-like focusing in a system of concentric rigid rings. Although the size of experimental system was reduced (to 7 rings), the main properties of the Bessel beam formation were demonstrated: the elongated foci along the symmetry axis, the Bessel-like distributions of the field in axial cross-sections, and the expected dependence of the focal distance with the frequency. The Bessel beam formation can be substantially improved, as follows from numerical calculations with larger number of rings.

For technical applications, the Bessel beam formation can be modified and improved, according to specific needs, by some means as for example: ($i$) the use of not toroidal rings, but rather the ones with more sophisticated shapes, which would favour the converging part of the diffracted wave (i.e. to convert into Bessel beams nearly all radiation (note that the rings can converge as maximum the half of initial radiation); ($ii$) the use of a ring structure of adequate thickness which would allow tailoring the longitudinal profile of the focus, according to the requirements; ($iii$) modifying the radii of the rings (making the radii incrementing not linearly) which is another parameter allowing the engineering of the focusing, and allowing to optimize the focal structure; finally ($iv$) by using multiple layers of rings at equidistant separations along the symmetry axis one could not only enhance the effect, but also introduce another possibility of tailoring the focal spot, as the interference from different planes will come into play.

\begin{acknowledgments}
The work was supported by Spanish Ministry of Science and Innovation and European Union FEDER through projects FIS2011-29731-C02-01 and -02, also MAT2009-09438, MTM2012-36740-C02-02 and UPV-PAID 2012/253.. VRG acknowledges financial support from the \lq\lq Pays de la Loire\rq\rq through the post-doctoral programme.
\end{acknowledgments}


%

\end{document}